\documentclass[12pt]{article}
\usepackage{amsmath}
\usepackage{amsmath}
\usepackage{graphicx}

\begin{document}

\title{\textbf{Two-level systems: exact solutions and underlying
pseudo-supersymmetry}}
\author{\textbf{V.V. Shamshutdinova$^{a}$, Boris F. Samsonov$^{a}$, and D.M.
Gitman$^{b}$}}
\date{$^{a}${\small \textit{Department of Quantum Field Theory, Tomsk State
University, 36 Lenin Ave., 634050 Tomsk, Russia }}\\
\small \textit{$^b$Institute of Physics, USP, Brazil}}
\maketitle

\begin{abstract}
Chains of first-order SUSY transformations
for the spin equation are studied in detail.
It is shown
that the transformation chains are related with a polynomial
pseudo-supersymmetry of the system. Simple determinant formulas for the
final Hamiltonian of a chain and for solutions of the spin equation are
derived. Applications are intended for a two-level atom in an
electromagnetic field with a possible time-dependence of the field
frequency. For a specific form of this dependence, the time oscillations of
the probability to populate the excited level disappear. Under certain
conditions this probability becomes a function tending monotonously to a
constant value which can exceed $1/2$.

PACS: 03.65.Fd, 11.30.Pb
\end{abstract}

\section{Introduction}

It is well-known that complex quantum systems with a discrete energy
spectrum (i.e., an atom in an external electric field) can be placed in a
special dynamical configuration in which only two stationary states are
essential. In those cases the description of a system's evolution does not
require a consideration of the entire Hilbert space. A good approximation is
achieved in case the whole Hilbert space is replaced by a two-dimensional
linear space. Systems of this kind are called two-level systems, whose
evolution is governed by a special system of two differential equations,
known as the ``spin equation''.

The spin equation arises in many areas of theoretical physics and thus finds
a wide range of applications, for instance, the semi-classical theory of
laser beams \cite{Nus73}, the absorption resonance and nuclear induction
experiments \cite{RabRaS54}, the behavior of a molecule in a cavity immersed
in electric or magnetic fields \cite{FeyVe57}. Particularly, we would like
to stress the importance of the spin equation in view of its possible
applications to quantum computations \cite{Qcomp}.

The first exact solution of the spin equation was obtained by Rabi \cite%
{Rabi}. This solution found an extensive physical application in the study
of various properties of two-level systems. This fact shows a great
importance of exactly solvable models involving the spin equation. In a
recent paper \cite{Levin}, some general properties of the spin equation were
studied. The authors of \cite{Levin}\ give an overview of the known results
and obtain new classes of exact solutions to this equation. Among the
methods of analysis of exactly solvable spin equations, the method of
intertwining operators \cite{Bagrov}--\cite{CzJP} plays a special role,
since it reveals a new type of symmetry related with two-level systems \cite%
{Shamshutdinova,CzJP}, called the polynomial pseudo-supersymmetry.

In this respect, we would like to remind that the ideas and methods of
supersymmetry (SUSY) are recently coming into use in relativistic and
nonrelativistic quantum-mechanical problems. Supersymmetric quantum
mechanics (SUSY QM), originated from the simplest quantum field model in
connection with the problem of a spontaneous SUSY breaking \cite{Witten},
finds more and more extensive applications in theoretical physics (see,
e.g., \cite{JPhys}). It was discovered that diverse aspects of SUSY QM are
closely related with the method of Darboux transformations, well-known in
the soliton theory \cite{Soliton}. Numerous works are devoted to the
relation between SUSY QM and the technique of intertwining operators \cite%
{Cooper,Nieto}.

This article is a continuation of a previous works devoted to the study of
exact solutions of two-level systems and the peculiarities related to the
corresponding underlying supersymmetry \cite{Bagrov}--\cite{CzJP}. Our main
objective is to show that a repeated use of simple transformations
introduced in the above papers gives rise to a wide range of exactly
solvable interactions for the spin equation. It turns out that such a study
is convenient in the analysis of equations describing two-level systems
written in the form of a one-dimensional stationary Dirac equation with a
non-Hermitian Hamiltonian of a special form, in which time plays the role of
a spatial variable \cite{Bagrov}. In this respect, we can mention that
non-Hermitian interaction arises, in particular, in various problems of
field theory, statistical mechanics (see, e.g., \cite{Itzykson}) and nuclear
physics. In connection with the discovery of a large class of complex
potentials possessing the so-called $\mathcal{PT}$-symmetry (see, e.g., \cite%
{Cannata}), which often leads to Hamiltonians with a purely real spectrum,
recently an attempt was made once again to construct a complex extension of
quantum mechanics \cite{Bender1}. In this respect, it is worth mentioning
the paper \cite{QuasiH}, where the authors show that the use of
non-Hermitian (so-called quasi-Hermitian) operators does not contradict the
basic principles of quantum mechanics. It should also be noted that such a
generalization was not started from scratch, and, in fact, partially goes
back to classical works \cite{90302141}.

In this respect, the role of supersymmetry in quantum mechanics may increase %
\cite{Shamshutdinova}, since, as was recently shown, SUSY transformations
allow one to remove singular points from continuous spectra of non-Hermitian
Hamiltonians \cite{my2}, and to convert non-diagonalizable Hamiltonians into
diagonalizable ones \cite{my1}. This may enlarge the set of admissible
super-Hamiltonians, thus giving an opportunity to extend the theory by new
types of interaction.

In \cite{Shamshutdinova} it was shown how to construct matrix-differential
intertwining operators preserving the special form of a non-Hermitian Dirac
Hamiltonian, and in \cite{CzJP} some results were announced related to
transformation chains. The technique of simple (first-order in derivatives)
intertwining operators adapted to the special case of Rabi oscillations made
it possible to discover an interesting physical effect \cite{Shamshutdinova}%
. Namely, as distinct form a constant field frequency when the
probability to populate the excited level oscillates with time
(Rabi oscillations), there exist such types of time dependence of
the field frequency that this probability ceases to oscillate and
becomes a monotonously increasing function of time tending to a
value which may exceed $1/2$ (in particular, it may be equal to
$3/4$ \cite{Shamshutdinova}). This property gives the hope that
inverse population may be observed in an ensemble of two-level
atoms placed in such type of field, and they may exhibit lasing
properties. This result was obtained when a single Darboux
transformation was applied to a two-level atom.

One of the aims of the present work is a detailed analysis of the
above-mentioned transformation chains. In order to make this work
self-contained, we show first of all (see the following section) how the Schr%
\"{o}dinger equation for a two-level atom interacting in the rotating wave
approximation with the electric component of an electromagnetic field can be
reduced to a one-dimensional stationary Dirac equation with an effective
non-Hermitian Hamiltonian where time plays the role of a spatial variable.
This is accomplished for a general time-dependence of the field frequency.

The third section is devoted to supersymmetric constructions. It is shown
that the transformation chains introduced in \cite{Bagrov,Shamshutdinova}
are related with a polynomial pseudo-supersymmetry of the system. This fact
indicates the presence of this type of symmetry in a two-level atom.

The introductory part of the fourth section contains a brief review of
previous results which are required in the following sections. It is shown
that in the case of various factorization constants the use of formulas
similar to the Crum--Krein formulas \cite{Crum} for the Schr\"{o}dinger
equation permits us to express the potential, obtained as a result of an $n$%
-fold SUSY transformation, through $n$-order determinants of the
transformation functions, thus avoiding the use of recurrent formulas, which
require the knowledge of solutions at all the intermediary steps. We obtain
the form of transformation functions that preserves both the specific matrix
structure of the initial potential and its real-valued character.
 We also study
the case of coinciding factorization constants and obtain a sufficiently
simple realization of such transformation chains.

As a numerical application of our general scheme, in the final (fifth)
section we examine a two-fold transformation for a two-level atom in an
external electromagnetic field. We discover that in this case the
disappearance of oscillations in the time-dependence of the probability to
populate the excited level may also occur, i.e., at certain conditions this
probability may acquire a monotonous time-dependence. It is established
that, in contrast to the previously published results \cite{Shamshutdinova},
this effect emerges for two kinds of behavior of the detuning of the
external field frequency from the resonance value. In these cases, the
probability values to populate the excited level after a two-fold
transformation coincide with either the maximal or the minimal values of the
probability oscillations obtained as a result of a one-fold transformation.

\section{Two-level atom in external field}

\label{part1}

Since the results obtained in this paper are mainly applied to a two-level
atom in an external electromagnetic field, in this section we briefly review
some properties of this system which are required in the following sections.

In the non-relativistic approximation, the evolution of a state vector of a
two-level atom in an external electromagnetic field with the electric
component $\mathbf{E}\left( t\right) =\mathbf{e}E_{0}\cos \left[ \omega
\left( t\right) t\right] $, where $\mathbf{e}$ is a unit vector of the field
polarization, is described by the Schr\"{o}dinger equation%
\begin{equation}
i\hbar \dot{\psi}\left( \mathbf{r},t\right) =\left( H_{0}+H_{1}\right) \psi
\left( \mathbf{r},t\right) \,.  \label{1}
\end{equation}%
Here, $H_{0}$ is the conventional
Hamiltonian of an atom in the absence of an external
field, $H_{0}\psi _{n}(\mathbf{r})=\varepsilon _{n}\psi _{n}(\mathbf{r})$, $%
n=1,2;$ $H_{1}$ describes the interaction of the atom with the electric
field component in the dipole approximation,%
\begin{equation*}
H_{1}=\left| q\right| E_{0}\left( \mathbf{e}\cdot \mathbf{r}\right) \cos %
\left[ \omega \left( t\right) t\right] \,;
\end{equation*}%
$q<0$ is the charge of electron; the dot over a symbol stands for
differentiation with respect to time. We assume that the field frequency, $%
\omega =\omega \left( t\right) $, and thus also the detuning of the
frequency from the resonance value, $\omega _{21}=\omega _{2}-\omega _{1}$,
\begin{equation}
\delta \left( t\right) =\omega (t)-\omega _{21}\,,  \label{delta}
\end{equation}%
are functions of time. Decomposing the solution $\psi \left( \mathbf{r}%
,t\right) $ of equation (\ref{1}) in the basis $\psi _{n}\left( \mathbf{r}%
\right) \exp \left( -i\omega _{n}t\right) $, $\omega _{n}=\varepsilon
_{n}/\hbar $, $n=1,2,$ we obtain the well-known \cite{Orszag} set of
equations for the decomposition coefficients $C_{n}$, $n=1,2$. Introducing
the notation $A_{1,2}=\exp \left( \mp i\frac{\delta }{2}t\right) C_{1,2}$
and neglecting the term $\exp \left[ i(\omega _{21}+\omega (t))t\right] $,
rapidly oscillating in comparison with $\exp \left[ i\delta t\right] $ (the
so-called rotating wave approximation \cite{Orszag,Allen}), we arrive at the
following set of equations:
\begin{equation}
i\dot{A}_{1}-fA_{1}=\xi A_{2}\,,\quad i\dot{A}_{2}+fA_{2}=\xi A_{1}\,.
\label{2}
\end{equation}%
Here, $\xi =\frac{1}{2\hbar }E_{0}d_{21},$%
\begin{equation}
f=f(t)=\frac{1}{2}\frac{d}{dt}\left[ t\delta (t)\right] \text{\thinspace },
\label{f}
\end{equation}%
and $d_{21}$ is a matrix element of the dipole momentum operator, $%
d_{12}=d_{21}=|q|\langle \psi _{1}|\left( \mathbf{e\cdot r}\right) |\psi
_{2}\rangle $. If $\omega $ does not depend on time (consequently, $f=\frac{1%
}{2}\delta =\mathrm{const}$), then solving equations \eqref{2} with the
initial conditions $A_{2}=0$ and $A_{1}=1$ for $t=0$ we determine the
probability to detect the system at the moment $t$ in the excited state on
condition that at the initial time moment $t=0$ the system was at the ground
state:%
\begin{equation}
|A_{2}\left( t\right) |^{2}=\frac{\xi ^{2}}{2\Omega ^{2}}\left[ 1-\cos
\left( 2\Omega t\right) \right] ,\quad \Omega ^{2}=f^{2}+\xi ^{2}\,.
\label{Rabi}
\end{equation}%
This result is well-known in quantum optics \cite{Orszag,Allen}. Formula (%
\ref{Rabi}) describes the so-called Rabi oscillations, and the value equal
to $2\xi $, is known as the Rabi frequency.

Introducing the notation%
\begin{equation}
V_{0}(t)=i\sigma _{2}f_{0}\left( t\right) ,  \label{V0}
\end{equation}%
$\gamma =i\sigma _{1}$, $E=\xi $, $\Psi =\left( A_{1},A_{2}\right) ^{\top }$
(the symbol${}^{\top }$ stands for transposition, while $\sigma _{1,2,3}$
are the standard Pauli matrices) we rewrite the system of equations (\ref{2})
at $f=f_{0}$ in the matrix form%
\begin{equation}
h_{0}\Psi =E\Psi \,,\quad h_{0}=\gamma \frac{d}{dt}+V_{0}\left( t\right) \,.
\label{h0}
\end{equation}%
Equation (\ref{h0}) has the form of a one-dimensional stationary Dirac
system where time plays the role of a spatial variable and $V_{0}\left(
t\right) $ is a matrix-valued potential determined by the function $%
f_{0}\left( t\right) $, which we shall call the potential. Following the
well-established terminology, we call $h_{0}$ the Hamiltonian even though in
the current case it does not correspond to any quantum-mechanical system. By
construction, the parameters $f_{0}$ and $E$ are real.

\section{Polynomial pseudo-supersymmetry of a two-level system}


Let us assume that the solutions of equation (\ref{h0}) are known and it is
necessary to find solutions of this equation with another potential,%
\begin{equation}
h_{1}\Phi =E\Phi \,,\quad h_{1}=\gamma \frac{d}{dt}+V_{1}\left( t\right) \,,
\label{h1}
\end{equation}%
where $V_{1}(t)=i\sigma _{2}f_{1}\left( t\right) $. This problem can be
solved by finding such an operator $L_{0,1}$ that obeys the following
operator equality (intertwining relation):
\begin{equation}
L_{0,1}h_{0}=h_{1}L_{0,1}\,.  \label{intertwi}
\end{equation}%
In this case, solutions of equation (\ref{h1}) can be found by applying the
operator $L_{0,1}$ to solutions of the initial equation (\ref{h0}), $\Phi
=L_{0,1}\Psi $.

The intertwining relation (\ref{intertwi}) provides the basis for the
general concept of transformation operators (see, e.g., \cite{Levitan}),
and, in particular, operators of Darboux transformations \cite{TMP}. At the
same time, the existence of the intertwining operator $L_{0,1}$ allows one
to construct a (generally polynomial) supersymmetry algebra \cite{TMP}
related to equation (\ref{h0}), thus revealing its internal supersymmetric
nature.

Usually in optical problems, as distinct from quantum-mechanical problems, the
introduction of a Hilbert space is unnecessary. In this case, symmetry
operators are defined on the space of solutions of the corresponding
equation without introducing any inner product. In the case of a two-level
atom, the equation in question is the differential equation (\ref{h0}). For
supersymmetric constructions, we shall need the notion of operator
conjugation, which will be introduced in a formal way. The operation of
formal (Laplace) conjugation obeys the standard rules $(AB)^{\dag }=B^{\dag
}A^{\dag }$, $(d/dt)^{\dag }=-d/dt$ and corresponds to the transposition of
a matrix accompanied by the complex conjugation of its elements. The
operator $A$ is called Hermitian if $A^{\dag }=A$. It is easy to seen from (%
\ref{h0}) that $h_{0}$ is non-Hermitian. Therefore, along with equation (%
\ref{h0}) there also exists its adjoint form%
\begin{equation}
h_{0}^{\dag }\widetilde{\Psi }=E\widetilde{\Psi }\,.  \label{h0d}
\end{equation}%
Another relation that we shall need is obtained by the conjugation of the
intertwining relation (\ref{intertwi})%
\begin{equation}
L_{0,1}^{\dag }h_{1}^{\dag }=h_{0}^{\dag }L_{0,1}^{\dag }\,.  \label{intpl}
\end{equation}%
The operator $L_{0,1}$, obviously transforms solutions of equation (\ref{h0}%
) into solutions of equation (\ref{h1}), while the operator $L_{0,1}^{\dag }$
realizes the backward transformation, i.e., a transformation from solutions
of the equation
\begin{equation*}
h_{1}^{\dag }\widetilde{\Phi }=E\widetilde{\Phi }\,,\qquad \widetilde{\Phi }%
=L_{0,1}^{\dag }\widetilde{\Psi }
\end{equation*}%
into solutions of equation (\ref{h0d}). We shall also assume that the
functions $\Psi $, $\widetilde{\Psi }$ and $\Phi $, $\widetilde{\Phi }$, as
well as the operators $h_{0}$, $h_{0}^{\dag }$ and $h_{1}$, $h_{1}^{\dag }$
are interrelated by an operator $J$%
\begin{equation*}
\widetilde{\Psi }=J\Psi \,,\qquad \widetilde{\Phi }=J\Phi \,,
\end{equation*}%
which is assumed to have the properties%
\begin{equation}
h_{0,1}^{\dag }=Jh_{0,1}J\,,\qquad J^{2}=\pm 1\,,\qquad J^{\dag }=\pm J\,.
\label{h0sigma}
\end{equation}%
From these equations it follows that a superposition of conjugation and the
operation $J$ is a pseudo-conjugation, while the operators $h_{0}$ and $%
h_{1} $ are pseudo-Hermitian \cite{Mostafazadeh}.

Relations (\ref{intpl}) and (\ref{h0sigma}) yield the equality%
\begin{equation*}
JL_{0,1}^{\dag }Jh_{1}=h_{0}JL_{0,1}^{\dag }J\,.
\end{equation*}%
Thus, the operator $JL_{0,1}^{\dag }J$ transforms solutions of the equation
with the Hamiltonian $h_{1}$ into solutions corresponding to the Hamiltonian
$h_{0},$ while the superposition $JL_{0,1}^{\dag }JL_{0,1}$ transforms
solutions of equation (\ref{h0}) into solutions of the same equation, and,
therefore, it is a symmetry operator of this equation. In exactly the same
manner, the operator $L_{0,1}JL_{0,1}^{\dag }J$ is a symmetry operator of
the equation with the Hamiltonian $h_{1}$. Since both these operators are
differential, one can expect that the indicated superpositions are
polynomials of the corresponding Hamiltonians, a property which we shall
demonstrate.

It is easy to see that in the case of Hamiltonian (\ref{h0}) we have $%
J=\sigma _{x}$. In \cite{Shamshutdinova} the authors constructed
matrix-differential intertwining operators preserving the specific form (\ref%
{h0}) of the non-Hermitian Dirac Hamiltonian $h_{0}$ and expressed the
symmetry operators $JL_{0,1}^{+}JL_{0,1}$ and $L_{0,1}JL_{0,1}^{+}J$ of
equations (\ref{h0}) and (\ref{h1}) in terms of the corresponding
Hamiltonians:%
\begin{equation}
JL_{0,1}^{+}JL_{0,1}=h_{0}^{2}-\Lambda _{1}^{2}\,,\quad
L_{0,1}JL_{0,1}^{+}J=h_{1}^{2}-\Lambda _{1}^{2}\,.  \label{JLJ}
\end{equation}%
The constant matrix $\Lambda _{1}=\mathrm{diag}(\lambda _{1},-\lambda _{1})$
in (\ref{JLJ}) is called the (matrix) factorization constant ($\lambda _{1}$
is also called the factorization constant). Formulas (\ref{JLJ}) present a
generalization of the factorization properties of transformation operators
that take place in the case of Hermitian one-component Hamiltonians \cite%
{Mielnik}.

The presence of a polynomial pseudo-supersymmetry in the system
under consideration is related with the possibility of a repeated
use of the above one-fold transformation, i.e., with the
realization of transformation chains. Let a sequence of
transformation operators $L_{0,1}$, $L_{1,2}$, \ldots ,
$L_{n-1,n}$ intertwines non-Hermitian Hamiltonians $h_{0}$,
$h_{1}$,
\ldots , $h_{n}$ and all the operators $h_{k}$ have potentials of the form $%
V_{k}\left( t\right) =i\sigma _{2}f_{k}\left( t\right) $. In this case,
every $h_{k}$ has the property%
\begin{equation}
h_{k}^{+}=Jh_{k}J\,,\quad k=0,\ldots ,n\,.  \label{hkp}
\end{equation}%
If we are interested only in the resulting action of the chain, we can
introduce the operator $L_{0,n}=L_{n-1,n}\ldots L_{1,2}L_{0,1}$. Due to the
intertwining relations which hold for each separate operator $L_{k-1,k}$
\begin{equation*}
L_{k-1,k}h_{k-1}=h_{k}L_{k-1,k}\,,\quad k=1,\ldots ,n,
\end{equation*}%
the operator $L_{0,n}$ obeys the relation%
\begin{equation}
L_{0,n}h_{0}=h_{n}L_{0,n}\,.  \label{intN}
\end{equation}%
Conjugating (\ref{intN}) and using (\ref{hkp}), we obtain%
\begin{equation}
JL_{0,n}^{+}Jh_{n}=h_{0}JL_{0,n}^{+}J\,.  \label{intJp}
\end{equation}%
Moreover, for every operator $h_{k}$ there holds the following
factorization:
\begin{equation*}
JL_{k-1,k}^{+}JL_{k-1,k}=h_{k-1}^{2}-\Lambda _{k}^{2}\,,\quad k=1,\ldots ,n
\end{equation*}%
with $\Lambda _{k}=\mathrm{diag}(\lambda _{k},-\lambda _{k})$.

Let us consider the superposition%
\begin{equation*}
JL_{0,n}^{+}JL_{0,n}=JL_{0,1}^{+}L_{1,2}^{+}\ldots
L_{n-1,n}^{+}JL_{n-1,n}\ldots L_{1,2}L_{0,1}\,.
\end{equation*}%
A subsequent use of separate factorization relations of type (\ref{JLJ}) and
intertwining relations of type (\ref{intertwi}) yields%
\begin{equation}
JL_{0,n}^{+}JL_{0,n}=(h_{0}^{2}-\Lambda _{1}^{2})(h_{0}^{2}-\Lambda
_{2}^{2})\ldots (h_{0}^{2}-\Lambda _{n}^{2})\,.  \label{fac0}
\end{equation}%
In a similar way, we obtain the factorization of the same polynomial of the
operator $h_{n}$
\begin{equation}
L_{0,n}JL_{0,n}^{+}J=(h_{n}^{2}-\Lambda _{1}^{2})(h_{n}^{2}-\Lambda
_{2}^{2})\ldots (h_{n}^{2}-\Lambda _{n}^{2})\,.  \label{facN}
\end{equation}

Let us introduce the following $4\times 4$ matrices:%
\begin{equation*}
H=%
\begin{pmatrix}
h_{0} & 0 \\
0 & h_{n}%
\end{pmatrix}%
\,,\quad Q_{1}=%
\begin{pmatrix}
0 & 0 \\
L_{0,n} & 0%
\end{pmatrix}%
\,,\quad Q_{2}=%
\begin{pmatrix}
0 & JL_{0,n}^{+}J \\
0 & 0%
\end{pmatrix}%
.
\end{equation*}%
It is easy to see that the intertwining relations (\ref{intN}) and (\ref%
{intJp}) are equivalent to the commutation relations%
\begin{equation}
\lbrack Q_{1},H]=[Q_{2},H]=0,  \label{comm}
\end{equation}%
while formulas (\ref{fac0}) and (\ref{facN}) can be rewritten in the form
\begin{equation}
Q_{1}Q_{2}+Q_{2}Q_{1}=(H^{2}-\Gamma _{1}^{2})(H^{2}-\Gamma _{2}^{2})\ldots
(H^{2}-\Gamma _{n}^{2})  \label{antic}
\end{equation}%
where $\Gamma _{k}=\mathrm{diag}(\Lambda _{k},\Lambda _{k})$. It is also
obvious that the operators $Q_{1}$ and $Q_{2}$ are nilpotent. Formulas (\ref%
{comm}) and (\ref{antic}) demonstrate that the operators $H$, $Q_{1}$ and $%
Q_{2}$ close a polynomial superalgebra thus revealing a hidden polynomial
supersymmetry of the system described by the Hamiltonian $h_{0}$. For the
first time, superalgebras related with non-Hermitian Hamiltonians were
introduced in \cite{Mostafazadeh}.
From this standpoint, a
polynomial pseudo-supersymmetry
can be associated with a
two-level system.

\section{Chains of transformations}

\subsection{General remarks}

Supersymmetric constructions of the previous section are valid if there
exist transformation operators $L_{0,1}$,\ldots ,$L_{n-1,n}$ preserving the
special form (\ref{h0}) of a non-Hermitian Dirac Hamiltonian at each step of
transformation where the potential matrix has the form (\ref{V0}). In \cite%
{Shamshutdinova} the authors constructed one-fold matrix-differential
intertwining operator $L_{0,1}$ having the necessary properties. In this
section, we shall sketch the results of \cite{Shamshutdinova} that we shall
need in the following sections.

Let us assume that we know $n$ matrix-valued solutions of the equation%
\begin{equation}
\left( \gamma \partial _{t}+V_{0}\right) \mathcal{U}_{k}=\mathcal{U}%
_{k}\Lambda _{k},\quad k=1,\ldots ,n\,.  \label{Lambdak}
\end{equation}%
This equation always has a solution,%
\begin{equation}
\mathcal{U}_{k}=\left(
\begin{array}{cc}
u_{k} & u_{k} \\
\tilde{u}_{k} & -\tilde{u}_{k}%
\end{array}%
\right) ,  \label{Uk}
\end{equation}%
corresponding to the factorization matrix $\Lambda _{k}=\mathrm{diag}\left(
iR_{k},-iR_{k}\right) $. The functions $u_{k}$, $\tilde{u}_{k}$ have the
same absolute value, and the sum of their phases is a constant, which we
denote by $C;$ see eqs. (\ref{u1}), (\ref{u2})). We choose the matrix-valued
function $\mathcal{U}_{1}$ as a transformation function for the first
transformation step. The functions $u_{1}$, $\tilde{u}_{1}$ defining the
matrix $\mathcal{U}_{1}$ are expressed in terms of a real function $q_{1},$
\begin{equation}
u_{1}=\frac{1-iq_{1}}{\sqrt{1+q_{1}^{2}}}\exp \left[ R_{1}\int \frac{%
1-q_{1}^{2}}{1+q_{1}^{2}}dt\right] e^{i\frac{C}{2}},  \label{u1}
\end{equation}%
\begin{equation}
\tilde{u}_{1}=\frac{1+iq_{1}}{\sqrt{1+q_{1}^{2}}}\exp \left[ R_{1}\int \frac{%
1-q_{1}^{2}}{1+q_{1}^{2}}dt\right] e^{i\frac{C}{2}},  \label{u2}
\end{equation}%
which, in turn, should be found with the help of a real solution $\chi _{1}$
of the second-order equation%
\begin{equation}
\ddot{\chi}_{1}+\left[ f_{0}^{2}+\frac{1}{2}\frac{d^{2}}{dt^{2}}\left( \ln
f_{0}\right) -\left( \frac{\ddot{f}_{0}}{2f_{0}}-R_{1}\right) ^{2}\right]
\chi _{1}=0  \label{eqchi}
\end{equation}%
as follows:
\begin{equation}
q_{1}=\frac{R_{1}}{f_{0}}-\frac{\dot{f}_{0}}{2f_{0}^{2}}-\frac{\dot{\chi}_{1}%
}{f_{0}\chi _{1}}\,.  \label{q}
\end{equation}%
The transformation operator of the first transformation step $L_{0,1}$ and a
new potential $V_{1}$ have the form%
\begin{align}
& L_{0,1}=\frac{d}{dt}-W_{1},\quad W_{1}=\dot{\mathcal{U}}_{1}\mathcal{U}%
_{1}^{-1}\,,  \label{L01} \\
& V_{1}=V_{0}+\left[ \gamma ,D_{1}\right] ,\quad D_{1}=W_{1}\,.  \label{V01}
\end{align}%
The potential matrix $V_{1}$ (\ref{V01}) has the form (\ref{V0}), where $%
f_{0}$ should be replaced by $f_{1}$. The function $f_{1}$ and the matrix $%
W_{1}$ defining the operator $L_{0,1}$ are expressed in terms of the same
function $q_{1},$
\begin{align}
& f_{1}=\frac{4R_{1}q_{1}}{1+q_{1}^{2}}-f_{0}\,,  \label{deltafq} \\
& W=\mathrm{diag}\left( w_{1},w_{2}\right) \,,\quad w_{1}=-if_{0}+R_{1}\frac{%
\left( 1+iq_{1}\right) ^{2}}{1+q_{1}^{2}}\,,\quad w_{2}=w_{1}^{\ast }\,.
\label{Wd}
\end{align}

If we choose $\lambda _{1}$ to be purely imaginary, the parameter $R_{1}$
becomes real and the differential equation (\ref{eqchi}) has real
coefficients. Due to the fact that an equation with real coefficients always
has real solutions, we conclude that for any real-valued function $f_{0}(t)$
one can obtain a real-valued function $f_{1}(t)$ thus realizing a
transformation with a real-valued resulting potential.

We now observe that the functions $\mathcal{V}_{k}=L_{0,1}\mathcal{U}_{k}$, $%
k=2,\ldots ,n$ are matrix-valued solutions of equation (\ref{Lambdak}) with
potential (\ref{V01}). At the next step of transformation, we choose $%
\mathcal{V}_{2}$ as a transformation function, thus obtaining the operator $%
L_{1,2}$ and potential $V_{2}$
\begin{align}
& L_{1,2}=\frac{d}{dt}-W_{2},\quad W_{2}=\dot{\mathcal{V}}_{2}\mathcal{V}%
_{2}^{-1}\,,  \label{L12} \\
& V_{2}=V_{1}+\left[ \gamma ,W_{2}\right] =V_{0}+\left[ \gamma ,D_{2}\right]
,\quad D_{2}=D_{1}+W_{2}\,.  \notag
\end{align}%
The operator $L_{0,2}=L_{1,2}L_{0,1}$ is a differential operator of second
order with matrix-valued coefficients, intertwining the Hamiltonians $h_{0}$
and $h_{2}=\gamma \partial _{t}+V_{2}$, $L_{0,2}h_{0}=h_{2}L_{0,2}$.

Continuing this process, one can construct the $n$-th order operator%
\begin{equation}
L_{0,n}=L_{n-1,n}\ldots L_{1,2}L_{0,1}\,,  \label{L0n}
\end{equation}%
which transforms the Hamiltonian $h_{0}$ into the Hamiltonian $h_{n}$ with
the potential%
\begin{equation}
V_{n}=V_{n-1}+[\gamma ,W_{n}]=V_{0}+\left[ \gamma ,D_{n}\right] ,\quad
D_{n}=D_{n-1}+W_{n}\,,\quad D_{0}=0\,,  \label{V0n}
\end{equation}%
where
\begin{equation}
W_{n}=\dot{\mathcal{Y}}_{n}\mathcal{Y}_{n}^{-1},\quad \mathcal{Y}%
_{n}=L_{0,n-1}\mathcal{U}_{n}\,.  \label{Yn}
\end{equation}

The derivation of the transformed potential $V_{n}$ by the recurrent
formulas (\ref{V0n}) and (\ref{Yn}) involves a calculation of all the
intermediary potentials and transformation functions, which requires a large
amount of computational work. Below, we will show how to overcome this
difficulty in the case when all the factorization constants $R_{1}$, \ldots
, $R_{n}$ are different from each other. For coinciding factorization
constants, we will show that equation (\ref{eqchi}) may be solved in
quadratures for any initial potential $f_{0}$ which also yields an essential
simplification of the method.

\subsection{Transformations with coinciding factorization constants}

\label{R1R2}

Let us consider a chain of two transformations. In order to obtain the
potential $f_{2},$ we have to make the replacement $R_{1}\rightarrow R_{2}$,
$q_{1}\rightarrow q_{2}$, $f_{1}\rightarrow f_{2}$, $f_{0}\rightarrow f_{1}$
in (\ref{deltafq}), and then $f_{1}$ in the obtained equation should be
substituted by using the same equation (\ref{deltafq}). This yields%
\begin{equation}
f_{2}=f_{0}-\frac{4R_{1}q_{1}}{1+q_{1}^{2}}+\frac{4R_{2}q_{2}}{1+q_{2}^{2}}%
\;.  \label{f2}
\end{equation}%
In (\ref{f2}) the function $q_{2}$ is expressed through the solution $\chi
_{2}$ of equation (\ref{eqchi}) with $f_{0}=f_{1}$ and $R_{1}=R_{2}$ by
formula (\ref{q}), where the necessary replacements should be made. A
generalization of this procedure to the case of a chain of $n$
transformations is obvious. However, the form of both the potential and
solutions to equation (\ref{eqchi}) after each transformation step will
become more and more complicated. It is remarkable that, in case the
transformation chain is realized by coinciding factorization constants,
equation (\ref{eqchi}) can be integrated explicitly. This makes it possible
to get rid of this equation and express the new potential $f_{2}$ in terms
of the potential $f_{1}$ alone.

Note, first of all, that according to (\ref{f2}), for $R_{1}=R_{2}$ and $%
q_{2}=q_{1}$ or $q_{2}=q_{1}^{-1},$ the potential $f_{2}$ assumes the value
of the initial potential $f_{0}$, i.e., the second transformation step is
equivalent to a reverse transformation. The case $q_{2}=q_{1}$ corresponds
to a trivial transformation, for which all of the three potentials remain
the same, $f_{2}=f_{1}=f_{0}$, whereas for $q_{2}=q_{1}^{-1}$ one obtains a
nontrivial result. Therefore, for $R_{2}=R_{1}$ the inverse transformation
corresponds to $q_{2}=q_{1}^{-1}$. Then using formula (\ref{q}) for $q_{2},$
in which we have to put $q_{1}=q_{2}$, $f_{0}=f_{1}$, $\chi _{1}=\chi _{2},$
we obtain the following differential equation with respect to the function $%
\chi _{2}$:
\begin{equation}
q_{1}^{-1}=\frac{R_{1}}{f_{1}}-\frac{\dot{f}_{1}}{2f_{1}^{2}}-\frac{\dot{\chi%
}_{2}}{f_{1}\chi _{2}}  \label{xiq}
\end{equation}%
which can be easily integrated:%
\begin{equation}
\chi _{2}=\tilde{C}\exp \int \left( R_{1}-\frac{\dot{f}_{1}}{2f_{1}}-\frac{%
f_{1}}{q_{1}}\right) dt\,.  \label{chi2}
\end{equation}%
Here, $\tilde{C}$ is an integration constant. Equation (\ref{eqchi}) with
fixed values of $f_{0}$ and $R_{1},$ being a second-order differential
equation, has two linearly independent solutions. Using the fact that the
Wronskian of any two linearly independent solutions to equation (\ref{eqchi}%
) is constant \cite{Kamke}, we can obtain a solution of this equation, which
is linearly independent from (\ref{chi2}). This permits us to find the
potential $f_{2}$ different from $f_{0}$, i.e., to accomplish the second
nontrivial step of transformation, with the same value of the factorization
constant. To this end, let us define the function $\tilde{\chi _{2}}$ as the
second solution of equation (\ref{eqchi}) satisfying the condition $W\left(
\chi _{2},\tilde{\chi}_{2}\right) =1,$ where $W$ denotes the Wronskian. Then
the function $\tilde{\chi}_{2}$ is expressed in terms of ${\chi }_{2}$ as
follows:
\begin{equation}
\tilde{\chi}_{2}=\chi _{2}\int \chi _{2}^{-2}dt\,.  \label{tildechi2}
\end{equation}%
Thus, in order to obtain a real potential by a two-fold transformation with
coinciding factorization constants, it is now unnecessary to solve equation (%
\ref{eqchi}) at the second step of transformation, since the structure of
its solution leading to a nontrivial result is determined and expressed in
terms of quantities known as a result of the first transformation step.

Let us express the result of consecutive application of transformations with
coinciding factorization constants in the general form. A solution of
equation (\ref{eqchi}) subject to the substitution $f_{0}\rightarrow f_{n-1}$
and $R_{1}\rightarrow R_{n}$ corresponding to the inverse transformation
such that $f_{n}=f_{n-2}$ has the form
\begin{equation}
\chi _{n}=\tilde{C}\exp \int \left( R_{n-1}-\frac{\dot{f}_{n-1}}{2f_{n-1}}-%
\frac{f_{n-1}}{q_{n-1}}\right) dt\,,  \label{chi22}
\end{equation}%
where $R_{n}=R_{n-1}$. By analogy with (\ref{tildechi2}), we obtain the
second solution $\tilde{\chi}_{n}$ of this equation linearly independent
from (\ref{chi22}), corresponding to another nontrivial step of a many-fold
transformation. The new potential can be found by formula (\ref{deltafq})
subject to the substitution $f_{0}\rightarrow f_{n-1}$ and $q_{1}\rightarrow
q_{n}$, where%
\begin{equation}
q_{n}=\frac{R_{n}}{f_{n-1}}-\frac{\dot{f}_{n-1}}{2f_{n-1}^{2}}-\frac{\dot{%
\tilde{\chi _{n}}}}{f_{n-1}\tilde{\chi}_{n}}\,.  \label{qn}
\end{equation}%
Thus, the use of transformation chains with coinciding factorization
constants allows one to easily obtain new exactly solvable potentials.

\subsection{Transformations with different factorization constants}

As we have already mentioned,
the explicit solution of equation (\ref{eqchi}) becomes
problematic in the case of a many-fold transformation. The integration of
this equation can be avoided by using formulas similar to the Crum--Krein
formulas \cite{Crum}, well-known for the Schr\"{o}dinger equation, which
allow one to express the potential after an $n$-fold transformation in terms
of determinants of transformation functions. This question was considered in
detail for a matrix Schr\"{o}dinger equation \cite{JourPhys} and a Dirac
equation \cite{Nieto} with Hermitian Hamiltonians. Below, we will
investigate the peculiarities arising during the use of these formulas for
the Dirac equation (\ref{h0}) with the effective non-Hermitian Hamiltonian $%
h_{0}$ and transformation functions of a special type, which do not change
the form of a Hamiltonian after a transformation and obey conditions
necessary for the transformed potential to remain a real value.

We start by reminding some results concerning chains of transformations for
the Dirac equation obtained in \cite{Nieto,JourPhys}. According to these
articles, the entries $d_{ij}^{n}$, $i,j=1,2$ of the matrix $D_{n}$ in (\ref%
{V0n}) are expressed in terms of the transformation functions $\mathcal{U}%
_{k}$, $k=1,\ldots ,n$ as follows:
\begin{equation*}
d_{ij}^{n}=\frac{|P_{ij}\left( \mathcal{U}_{1},\ldots ,\mathcal{U}%
_{n}\right) |}{|P\left( \mathcal{U}_{1},\ldots ,\mathcal{U}_{n}\right) |}\,.
\end{equation*}%
Here, $P\left( \mathcal{U}_{1},\ldots ,\mathcal{U}_{n}\right) $ is the
Wronsky matrix constructed from the matrix-valued functions $\mathcal{U}_{k}$%
, $k=1,\ldots ,n$,
\begin{equation}
P\left( \mathcal{U}_{1},\ldots ,\mathcal{U}_{n}\right) =\left(
\begin{array}{cccc}
\mathcal{U}_{1} & \mathcal{U}_{2} & \ldots & \mathcal{U}_{n} \\
\mathcal{U}_{1}{}^{\prime } & \mathcal{U}_{2}{}^{\prime } & \ldots &
\mathcal{U}_{n}{}^{\prime } \\
\ldots & \ldots & \ldots & \ldots \\
\mathcal{U}_{1}^{\left( n-1\right) } & \mathcal{U}_{2}^{\left( n-1\right) }
& \ldots & \mathcal{U}_{n}^{\left( n-1\right) }%
\end{array}%
\right) ;  \label{P}
\end{equation}%
the matrix $P_{ij}\left( \mathcal{U}_{1},\ldots ,\mathcal{U}_{n}\right) $
has a structure similar to that of $P\left( \mathcal{U}_{1},\ldots ,\mathcal{%
U}_{n}\right) $ with the exception that the final row consisting of matrices
$\mathcal{U}_{k}^{\left( n-1\right) }$ is replaced by matrices $\mathcal{U}%
_{k}^{ij}$, $k=1,\ldots ,n$:
\begin{equation*}
P_{ij}\left( \mathcal{U}_{1},\ldots ,\mathcal{U}_{n}\right) =\left(
\begin{array}{cccc}
\mathcal{U}_{1} & \mathcal{U}_{2} & \ldots & \mathcal{U}_{n} \\
\mathcal{U}_{1}{}^{\prime } & \mathcal{U}_{2}{}^{\prime } & \ldots &
\mathcal{U}_{n}{}^{\prime } \\
\ldots & \ldots & \ldots & \ldots \\
\mathcal{U}_{1}^{\left( n-2\right) } & \mathcal{U}_{2}^{\left( n-2\right) }
& \ldots & \mathcal{U}_{n}^{\left( n-2\right) } \\
\mathcal{U}_{1}^{ij} & \mathcal{U}_{2}^{ij} & \ldots & \mathcal{U}_{n}^{ij}%
\end{array}%
\right) .
\end{equation*}%
The matrices $\mathcal{U}_{k}^{ij}$, $i,j=1,2$ follow from $\mathcal{U}%
_{k}^{\left( n-1\right) }$ by the replacement of the $j$-th row by the $i$%
-th row of the matrix $\mathcal{U}_{k}^{\left( n\right) }$.

In \cite{JourPhys} it was shown that the action of the operator $L_{0,n}$ (%
\ref{L0n}) on the function $\Psi _{E}=\left( \psi _{1E},\psi _{2E}\right)
^{\top }$ yields a vector $\Phi _{E}=\left( \phi _{1E},\phi _{2E}\right)
^{\top }$ with the entries%
\begin{equation}
\phi _{jE}=\frac{|P_{jE}\left( \mathcal{U}_{1},\ldots ,\mathcal{U}%
_{n}\right) |}{|P\left( \mathcal{U}_{1},\ldots ,\mathcal{U}_{n}\right) |}%
\;,\quad j=1,2\;.  \label{phij}
\end{equation}%
In (\ref{phij}) the matrix $P_{jE}\left( \mathcal{U}_{1},\ldots ,\mathcal{U}%
_{n}\right) $ has the form
\begin{equation}
P_{jE}\left( \mathcal{U}_{1},\ldots ,\mathcal{U}_{n}\right) =\left(
\begin{array}{ccccc}
\mathcal{U}_{1} & \mathcal{U}_{2} & \ldots & \mathcal{U}_{n} & \Psi _{E} \\
\mathcal{U}_{1}{}^{\prime } & \mathcal{U}_{2}{}^{\prime } & \ldots &
\mathcal{U}_{n}{}^{\prime } & \Psi _{E}^{\prime } \\
\ldots & \ldots & \ldots & \ldots & \ldots \\
\mathcal{U}_{1}^{\left( n-1\right) } & \mathcal{U}_{2}^{\left( n-1\right) }
& \ldots & \mathcal{U}_{n}^{\left( n-1\right) } & \Psi _{E}^{\left(
n-1\right) } \\
\left( \mathcal{U}_{1}^{j}\right) ^{\left( n\right) } & \left( \mathcal{U}%
_{2}^{j}\right) ^{\left( n\right) } & \ldots & \left( \mathcal{U}%
_{n}^{j}\right) ^{\left( n\right) } & \psi _{jE}^{\left( n\right) }%
\end{array}%
\right) ,  \label{PjE}
\end{equation}%
where $\mathcal{U}_{k}^{j}$ is the $j$-th row of the matrix $\mathcal{U}_{k}$%
. These formulas are obtained as a closure of the recursion scheme
defined by equations (\ref{V0n}) and (\ref{Yn}). This algorithm
will preserve the particular form of potential (\ref{V0}) provided
that the transformation operators $L_{k-1,k}$ preserve the form
(\ref{Uk}) of the transformation function at every transformation
step, a property which we will now demonstrate by induction.

Let the transformation function $\mathcal{U}_{1}$ of the first
transformation step have the required form \eqref{Uk}. We also assume that
this property takes place at the $(k-1)$-st transformation step:%
\begin{equation}
\mathcal{V}_{j}=L_{0,k-2}\mathcal{U}_{j}=\left(
\begin{array}{cc}
v_{j} & v_{j} \\
\tilde{v}_{j} & -\tilde{v}_{j}%
\end{array}%
\right) ,\quad j=k-1,k,\ldots ,n\,.  \label{nuk}
\end{equation}%
Then the transformation function at the $k$-th step is expressed as follows:%
\begin{equation}
\mathcal{Y}_{k}=L_{k-2,k-1}\mathcal{V}_{k},\quad L_{k-2,k-1}=\left( \frac{d}{%
dt}-\dot{\mathcal{V}}_{k-1}\mathcal{V}_{k-1}^{-1}\right) .  \label{yk}
\end{equation}%
Using (\ref{nuk}), we find, just as we expected, that (\ref{yk}) implies%
\begin{equation*}
\mathcal{Y}_{k}=\left(
\begin{array}{cc}
\mathrm{y}_{k} & \mathrm{y}_{k} \\
\mathrm{\tilde{y}}_{k} & -\mathrm{\tilde{y}}_{k}%
\end{array}%
\right) ,\quad \mathrm{y}_{k}=\dot{v}_{k}-\frac{\dot{v}_{k-1}}{v_{k-1}}%
\,v_{k}\,.
\end{equation*}

It has been noted that for a purely imaginary parameters $\Lambda _{k}$ the
condition necessary for the transformed potential to be a real value is the
condition that the absolute values of the elements of the matrix-valued
transformation function $\mathcal{U}$ be equal. One can see that the set of
functions (\ref{Uk}) actually meets this condition. Therefore, these
functions allow one to realize a transformation chain which preserves both
the special form of the potential and the condition for the function $f_{k}$
to be real-valued.

We are now going to show that for the choice of transformation functions in
the form (\ref{Uk}) the matrices of $(2n)$-th order $P\left( \mathcal{U}%
_{1},\ldots ,\mathcal{U}_{n}\right) $, $P_{ij}\left( \mathcal{U}_{1},\ldots ,%
\mathcal{U}_{n}\right) $ assume a block form with one of the blocks being a
zero matrix. As a result, the determinants $|P\left( \mathcal{U}_{1},\ldots ,%
\mathcal{U}_{n}\right) |$, $|P_{ij}\left( \mathcal{U}_{1},\ldots ,\mathcal{U}%
_{n}\right) |$ are reduced to the product of two determinants of $n$-th
order. Let us consider in more detail how this happens for $n=2$, and then
present the result for an arbitrary $n$.

In what follows, the \textit{elements}, \textit{rows} and \textit{columns}
of a determinant are understood as the elements, rows and columns of the
corresponding matrix.

According to (\ref{P}), the determinant $|P\left( \mathcal{U}_{1},\ldots ,%
\mathcal{U}_{n}\right) |$ for $n=2$ has the form:
\begin{equation}
|P\left( \mathcal{U}_{1},\mathcal{U}_{2}\right) |=%
\begin{vmatrix}
u_{1} & u_{1} & u_{2} & u_{2} \\
\tilde{u}_{1} & -\tilde{u}_{1} & \tilde{u}_{2} & -\tilde{u}_{2} \\
u_{1}^{\prime } & u_{1}^{\prime } & u_{2}^{\prime } & u_{2}^{\prime } \\
{\tilde{u}_{1}}^{\prime } & {-\tilde{u}_{1}}^{\prime } & {\tilde{u}_{2}}%
^{\prime } & {-\tilde{u}_{2}}^{\prime }%
\end{vmatrix}%
.  \label{P2}
\end{equation}%
Subtracting from the second column the first one, and from the forth column
the third one, and transmuting the rows and columns, we reduce the
determinant (\ref{P2}) to the product of two second-order determinants:%
\begin{equation*}
|P\left( \mathcal{U}_{1},\mathcal{U}_{2}\right) |=%
\begin{vmatrix}
u_{1} & u_{2} & 0 & 0 \\
u_{1}^{\prime } & u_{2}^{\prime } & 0 & 0 \\
\tilde{u}_{1} & \tilde{u}_{2} & -2\tilde{u}_{1} & -2\tilde{u}_{2} \\
{\tilde{u}_{1}}^{\prime } & {\tilde{u}_{2}}^{\prime } & {-2\tilde{u}_{1}}%
^{\prime } & {-2\tilde{u}_{2}}^{\prime }%
\end{vmatrix}%
=%
\begin{vmatrix}
u_{1} & u_{2} \\
u_{1}^{\prime } & u_{2}^{\prime }%
\end{vmatrix}%
\left( -2\right) ^{2}%
\begin{vmatrix}
\tilde{u}_{1} & \tilde{u}_{2} \\
{\tilde{u}_{1}}^{\prime } & {\tilde{u}_{2}}^{\prime }%
\end{vmatrix}%
.
\end{equation*}%
It is clear that in general we have $|P\left( \mathcal{U}_{1},\ldots ,%
\mathcal{U}_{n}\right) |=\left(-2\right)^np_{1}p_{2}$, where%
\begin{equation*}
p_{1}=%
\begin{vmatrix}
u_{1} & u_{2} & \ldots & u_{n} \\
u_{1}^{\prime } & u_{2}^{\prime } & \ldots & u_{n}^{\prime } \\
\ldots & \ldots & \ldots & \ldots \\
u_{1}^{\left( n-2\right) } & u_{2}^{\left( n-2\right) } & \ldots &
u_{n}^{\left( n-2\right) } \\
u_{1}^{\left( n-1\right) } & u_{2}^{\left( n-1\right) } & \ldots &
u_{n}^{\left( n-1\right) }%
\end{vmatrix}%
,\quad p_{2}=%
\begin{vmatrix}
\tilde{u}_{1} & \tilde{u}_{2} & \ldots & \tilde{u}_{n} \\
\tilde{u}_{1}^{\prime } & \tilde{u}_{2}^{\prime } & \ldots & \tilde{u}%
_{n}^{\prime } \\
\ldots & \ldots & \ldots & \ldots \\
\tilde{u}_{1}^{\left( n-2\right) } & \tilde{u}_{2}^{\left( n-2\right) } &
\ldots & \tilde{u}_{n}^{\left( n-2\right) } \\
\tilde{u}_{1}^{\left( n-1\right) } & \tilde{u}_{2}^{\left( n-1\right) } &
\ldots & \tilde{u}_{n}^{\left( n-1\right) }%
\end{vmatrix}%
.
\end{equation*}%
Similar calculations allow one to simplify the determinants of the matrices $%
P_{11}\left( \mathcal{U}_{1},\ldots ,\mathcal{U}_{n}\right) $ and $%
P_{22}\left( \mathcal{U}_{1},\ldots ,\mathcal{U}_{n}\right) $:
\begin{equation*}
|P_{11}\left( \mathcal{U}_{1},\ldots
,\mathcal{U}_{n}\right)|=\left(-2\right)^nr_{1}p_{2}\,,\quad
|P_{22}\left( \mathcal{U}_{1},\ldots
,\mathcal{U}_{n}\right)|=\left(-2\right)^np_{1}r_{2}\,,
\end{equation*}%
where%
\begin{equation*}
r_{1}=%
\begin{vmatrix}
u_{1} & u_{2} & \hdots & u_{n} \\
u_{1}^{\prime } & u_{2}^{\prime } & \hdots & u_{n}^{\prime } \\
\ldots & \ldots & \ldots & \ldots \\
u_{1}^{\left( n-2\right) } & u_{2}^{\left( n-2\right) } & \hdots &
u_{n}^{\left( n-2\right) } \\
u_{1}^{\left( n\right) } & u_{2}^{\left( n\right) } & \hdots & u_{n}^{\left(
n\right) }%
\end{vmatrix}%
,\quad r_{2}=%
\begin{vmatrix}
\tilde{u}_{1} & \tilde{u}_{2} & \hdots & \tilde{u}_{n} \\
\tilde{u}_{1}^{\prime } & \tilde{u}_{2}^{\prime } & \hdots & \tilde{u}%
_{n}^{\prime } \\
\ldots & \ldots & \ldots & \ldots \\
\tilde{u}_{1}^{\left( n-2\right) } & \tilde{u}_{2}^{\left( n-2\right) } & %
\hdots & \tilde{u}_{n}^{\left( n-2\right) } \\
\tilde{u}_{1}^{\left( n\right) } & \tilde{u}_{2}^{\left( n\right) } & \hdots
& \tilde{u}_{n}^{\left( n\right) }%
\end{vmatrix}%
\,.
\end{equation*}%
The determinants of the matrices $P_{12}\left( \mathcal{U}_{1},\ldots ,\mathcal{U}%
_{n}\right) $ and $P_{21}\left( \mathcal{U}_{1},\ldots ,\mathcal{U}%
_{n}\right) $ for the transformation functions (\ref{Uk}) are equal to zero.
Threfore, the matrix $D_{n}$ from (\ref{V0n}) becomes $D_{n}=\mathrm{diag}%
\left( \frac{r_{1}}{p_{1}},\frac{r_{2}}{p_{2}}\right) $ and $V_{n}$ can be
written in the form
\begin{equation*}
V_{n}=V_{0}+\Delta V_{n},\quad \Delta V_{n}=i\sigma _{2}\Delta f_{n},\quad
\Delta f_{n}=f_{n}-f_{0}=-i\left[ \frac{r_{1}}{p_{1}}-\frac{r_{2}}{p_{2}}%
\right] .
\end{equation*}

Thus, we have shown that the use of the determinant technique for an $n$%
-fold transformation of the spin equation, which is similar to the
Crum--Krein approach to the Schr\"{o}dinger equation, permits us to reduce
the method to calculating only $n$-th order determinants.

\section{Transformations of the Rabi oscillations}

The Rabi oscillations occur at $f_{0}=\mathrm{const}$. In \cite%
{Shamshutdinova} a detailed analysis of solutions to the set of equations (%
\ref{h0}) constructed with the help of a one-fold transformation is given.
Here, we shall study the effect of a two-fold transformation.

Let us first consider the probability to populate
the excited level obtained
as a result of a two-fold transformation with coinciding factorization
constants $R_{2}=R_{1}$. For $f_{0}=\mathrm{const}$, equation (\ref{eqchi})
takes a very simple form:%
\begin{equation}
\ddot{\chi _{1}}+\varpi ^{2}\chi _{1}=0\,,\quad \varpi
^{2}=f_{0}^{2}-R_{1}^{2}.  \label{chi0}
\end{equation}%
Solutions of this equation have various properties depending on whether the
value of $\varpi ^{2}$ is positive, negative or zero. We have found \cite%
{Shamshutdinova} that the Rabi oscillations after a one-fold transformation
disappear for $\varpi =0$. In this case, the general solution of equation (%
\ref{chi0}) is a linear function of time, $\chi _{1}=At+B$, which in the
case $A=2Bf_{0}$ yields the following expression for the function $q_{1}$: $%
q_{1}=1-2/\left( 2f_{0}t+1\right) $. The potential after the one-fold
transformation reads%
\begin{equation*}
f_{1}(t)=f_{0}-\frac{4f_{0}}{1+4f_{0}^{2}t^{2}}\,.
\end{equation*}%
Let us make the next transformation with $R_{2}=R_{1}$. The function $q_{2}$
is found from (\ref{qn}) at $n=2$, where $\widetilde{\chi }_{2}$ is
determined by formulas (\ref{chi2})-(\ref{tildechi2}). Using (\ref{f2}), we
get the following expression for the transformed potential:
\begin{equation}
f_{2}\left( t\right) =\frac{%
f_{0}(-648(-1+4f_{0}^{2}t^{2})-243(1+4f_{0}^{2}t^{2})^{2}+(C-6f_{0}t(9+4f_{0}^{2}t^{2}))^{2})%
}{81(1+4f_{0}^{2}t^{2})^{2}+(C-6f_{0}t(-3+4f_{0}^{2}t^{2}))^{2}}\,,
\label{f22}
\end{equation}%
where $C$ is a constant.

Let us determine the form of the external electric field $E=E(t)=E_{0}\cos %
\left[ t\omega _{2}\left( t\right) \right] $ corresponding to the potential $%
f_{2}\left( t\right) $ (\ref{f22}). We may observe an interesting feature of
the behavior of $E(t)$ with time. From (\ref{delta}) and (\ref{f}), it
follows that the frequency $\omega _{2}\left( t\right) $ is expressed
through the function $f_{2}\left( t\right) $ as follows:
\begin{equation}
\omega _{2}\left( t\right) =\omega _{21}+\delta _{2}\left( t\right) ,\quad
\delta _{2}\left( t\right) =\frac{2}{t}\int_{0}^{t}f_{2}\left( \tau \right)
d\tau \,.  \label{d2t}
\end{equation}%
In general, the field $E(t)$ is an oscillating function with the amplitude $%
E_{0}$ (see Fig. 1)
except for a special time dependence of $\delta _{2}(t)$.
In particular, it may happen that $\delta _{2}(t)$,
as found from (\ref{d2t}),
is expressed in terms of
an inverse trigonometric function shifted by a constant $c_0$.
If $c_0+\omega _{21}=0$,
the function $\cos [t\omega _{2}(t)]$ may become polynomial.
For instance, substituting in (\ref%
{f22}) $C=0$ and $f_{0}=-1,$ we obtain for $\omega _{2}(t)$
\begin{equation*}
\omega _{2}(t)=\omega _{21}-2+\frac{4}{t}\arctan \left( \frac{8t^{3}-6t}{%
12t^{2}+3}\right) \,.
\end{equation*}%
Thus, for $\omega _{21}=2$ the field $E$ becomes the polynomial%
\begin{equation*}
E(t)=E_{0}\left( 1-\frac{5184(1+16t^{2}+48t^{4})}{%
(9+108t^{2}+48t^{4}+64t^{6})^{2}}+\frac{288(2+7t^{2}-4t^{4})}{%
9+108t^{2}+48t^{4}+64t^{6}}\right)
\end{equation*}%
plotted in Fig.2.
\begin{figure}[tbp]
\begin{center}
\includegraphics[
height=1.5in, width=2.5in ]{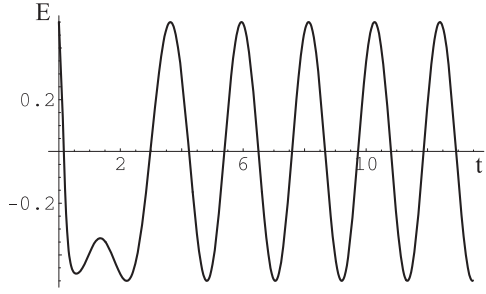}
\end{center}
\caption{The electric component of the external electromagnetic field $%
f_{0}=1$, $C=0$, $w_{21}=1$, $E_{0}=0.5$.}
\end{figure}

\begin{figure}[ptbp]
\begin{center}
\includegraphics[
height=1.5in, width=2.5in ]{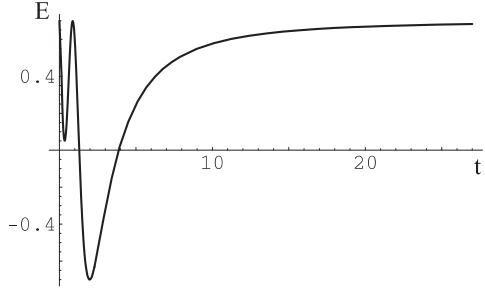}
\end{center}
\caption{The electric component of the external electromagnetic field after
a two-fold transformation with the parameter values $f_{0}=-1$, $C=0$, $%
w_{21}=2$, $E_{0}=0.7$.}
\end{figure}

Another interesting effect has been observed for a one-fold transformation %
\cite{Shamshutdinova}. It consists in the disappearance of oscillations in
the time-dependence of the excited level population. For a two-fold
transformation, there are additional possibilities for observing this
property.

Solutions of the initial problem for $f=f_{1}$ are known \cite%
{Shamshutdinova}. Having constructed the transformation operator by using (%
\ref{L01}) and (\ref{L12}), we find the solutions $A_{12}(t)$ and $A_{22}(t)$
of the set of equations after the second step of transformation.
Note that
the normalization of solutions changes as a result of the transformation $%
A_{12}^{2}+A_{22}^{2}=\left( \xi ^{2}+R_{1}^{2}\right) ^{2}$.
Imposing on
the solutions the initial conditions $A_{12}(0)=1$ and $A_{22}(0)=0,$ we
determine the probability to populate the excited level after a two-fold
transformation. The expression for the probability is rather involved and we
will not present it here. We merely note that this quantity oscillates with
time due to the presence of trigonometric functions.

Let us consider the case $C=0$ in more detail. It turns out that under the
condition
\begin{equation}
5f_{0}^{4}-10f_{0}^{2}\xi ^{2}+\xi ^{4}=0,  \label{condit}
\end{equation}%
the oscillations in the time dependence of the probability disappear and it
acquires a monotonous character:%
\begin{align*}
& P_{2}\left( t\right) =|A_{22}|^{2}= \\
& \frac{2000t^{2}\xi ^{2}(45(25+11\sqrt{5})+60(123+55\sqrt{5})t^{2}\xi
^{2}+32(1525+682\sqrt{5})t^{4}\xi ^{4})}{(5+\sqrt{5})^{5}(225+540(5+2\sqrt{5}%
)t^{2}\xi ^{2}+240(9+4\sqrt{5})t^{4}\xi ^{4}+64(85+38\sqrt{5})t^{6}\xi ^{6})}%
\,.
\end{align*}%
Therefore, according to (\ref{condit}), as distinct from one-fold
transformations, for two-fold transformations we can indicate two
possibilities, $f_{0}^{2}=\xi ^{2}\left( 1+\frac{2}{\sqrt{5}}\right) $ and $%
f_{0}^{2}=\xi ^{2}\left( 1-\frac{2}{\sqrt{5}}\right) $, which transform the
probability to populate the excited level from an oscillating function of
time to a monotonous one. The plots of probability in these cases are given
by Figs. 3 and 4 (thick lines).
\begin{figure}[tbp]
\begin{center}
\includegraphics[
height=1.5in,
width=2.5in
]{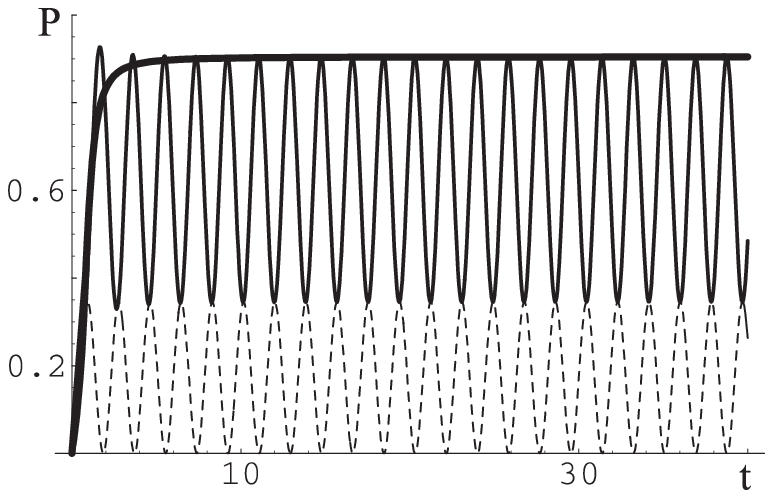}
\end{center}
\caption{The probability of population of the excited level before the
transformation, and after the first and second transformations for $%
f_{0}^{2}=\protect\xi ^{2}\left( 1+\frac{2\protect\sqrt{5}}{5}\right) $, $%
\protect\xi =1$.}
\end{figure}
\begin{figure}[tbp]
\begin{center}
\includegraphics[
height=1.5in,
width=2.5in
]{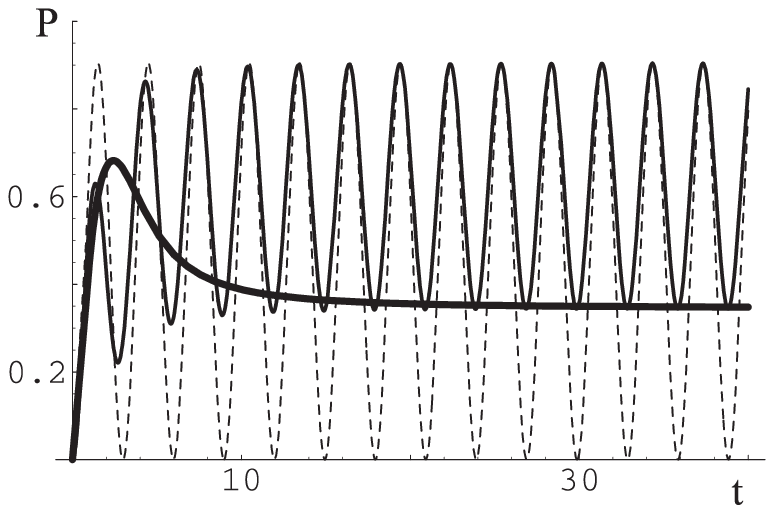}
\end{center}
\caption{The probability of population of the excited level before the
transformation, and after the first and second transformations for $%
f_{0}^{2}=\protect\xi ^{2}\left( 1-\frac{2\protect\sqrt{5}}{5}\right) $, $%
\protect\xi =1$.}
\end{figure}

We observe another interesting effect by plotting simultaneously the
probability to populate the excited level before the transformation and
after the first and second transformation steps. Fig. 5 presents the
probability plots for a particular value of the parameter $f_{0}^{2}=\frac{1%
}{3}\xi ^{2}$, which corresponds to the case when the probability of
population resulting from the one-fold transformation is described by a
monotonous function \cite{Shamshutdinova}
\begin{equation}
P_{1}(t)=|A_{12}|^{2}=\frac{3f_{0}^{2}t^{2}}{1+4f_{0}^{2}t^{2}}\,.
\label{P1}
\end{equation}%
The thick line presents the plot of probability (\ref{P1}). Its maximal
value coincides with the probability before the transformation, i.e., with
the probability for the usual Rabi oscillations (the dotted line). The thin
line presents the probability resulting from a chain of two transformations.
We see from Fig. 5 that the oscillations in this case are in antiphase with
the Rabi oscillations.
\begin{figure}[tbp]
\begin{center}
\includegraphics[
height=1.5in,
width=2.4in
]{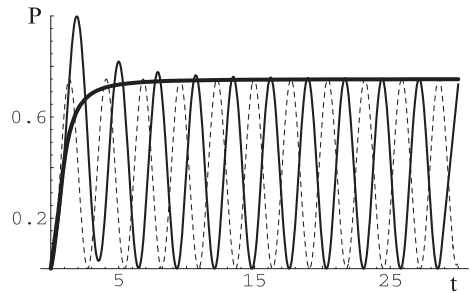}
\end{center}
\caption{The probability of population of the excited level before the
transformation, and after the first and second transformations for $%
f_{0}^{2}=\frac{1}{3}\protect\xi ^{2}$, $\protect\xi =1$.}
\end{figure}

Returning to Figs. 3 and 4, note that they also present the plots for $%
f_{0}^{2}=\xi ^{2}\left( 1+\frac{2}{\sqrt{5}}\right) $ and $f_{0}^{2}=\xi
^{2}\left( 1-\frac{2}{\sqrt{5}}\right)$, respectively.
For such values of $%
f_{0},$ the probability to populate the excited level after the two-fold
transformation becomes a monotonous function and for $t\rightarrow \infty $
the probability values after the second transformation (the thick line)
coincide with either the maxima or the minima of probability oscillations
resulting from the one-fold transformation (the thin line). The probability
before the transformation (the dotted line) oscillates in this case in phase
with the probability after the one-fold transformation.

Omitting the details, let us consider the behavior of the probability to
populate the excited level obtained as a result of a two-fold transformation
with different factorization constants. The transformed potential $f_{2}$ is
obtained in \cite{CzJP}
\begin{align}
& f_{2}=f_{0}-i\left( R_{2}^{2}-R_{1}^{2}\right) \frac{R_{2}\left[
B_{2}^{2}-\left( B_{2}^{\ast }\right) ^{2}\right] -R_{1}\left[
B_{1}^{2}-\left( B_{1}^{\ast }\right) ^{2}\right] }{%
|R_{2}B_{2}-R_{1}B_{1}|^{2}}\;,  \notag \\
&
B_{k}=\frac{u_{k}}{\tilde{u}_{k}}=\frac{1-iq_{k}}{1+iq_{k}}\;,\quad
k=1,2\,.  \label{f222}
\end{align}%
Using (\ref{phij}) and (\ref{PjE}), we construct the intertwining operator $%
L_{02}$ and then apply it to find solutions $A_{12}\left( t\right) $ and $%
A_{22}\left( t\right) $ of equation (\ref{h0}) with potential (\ref{f222}).
In terms of the functions $B_{k},$ the operator $L_{02}$ is written as
\begin{equation*}
L_{02}=\frac{d^{2}}{dt^{2}}+S_{1}\frac{d}{dt}+S_{2}\,,
\end{equation*}%
where
\begin{align*}
& S_{1}=\left(
\begin{array}{cc}
s_{1} & 0 \\
0 & s_{1}^{\ast }%
\end{array}%
\right) ,\quad s_{1}=\frac{R_{1}^{2}-R_{2}^{2}}{R_{2}B_{2}^{\ast
}-R_{1}B_{1}^{\ast }}\,, \\
& S_{2}=\left(
\begin{array}{cc}
s_{2} & 0 \\
0 & s_{2}^{\ast }%
\end{array}%
\right) ,\quad s_{2}=if_{0}^{\prime }+f_{0}^{2}+\frac{R_{1}R_{2}\left(
R_{2}B_{1}^{\ast }-R_{1}B_{2}^{\ast }\right) -if_{0}\left(
R_{2}^{2}-R_{1}^{2}\right) }{R_{2}B_{2}^{\ast }-R_{1}B_{1}^{\ast }}\,.
\end{align*}%
Let%
\begin{equation*}
q_{k}\left( t\right) =\frac{R_{k}\cos \varpi _{k}t+\frac{1}{2}\varpi
_{k}\sin \varpi _{k}t-f_{0}}{f_{0}\cos \varpi _{k}t-R_{k}},\;\varpi _{k}=2%
\sqrt{f_{0}^{2}-R_{k}^{2}},\;k=1,2.
\end{equation*}%
Acting by $L_{02}$ on solutions of the initial problem (\ref{h0}), we obtain
its solutions for $f_{0}=f_{2}$. Imposing on these solutions the initial
conditions $A_{12}\left( 0\right) =1$ and $A_{22}\left( 0\right) =0$, we
determine the probability to populate the excited level after a two-fold
transformation. In general, the behavior of probability has an oscillating
character. It is important to note that we now observe two kinds of
oscillations: small-amplitude and large-frequency (fast) oscillations occur
at the background of a large amplitude and small frequency (slow)
oscillations (see Fig. 6). Furthermore, at certain values of parameters, the
probability for fast oscillations changes within a narrow range (on Fig. 6
between $0.8$ and $0.95$) with the maximal value close to unity.
\begin{figure}[tbp]
\begin{center}
\includegraphics[
height=1.5in,
width=3in
]{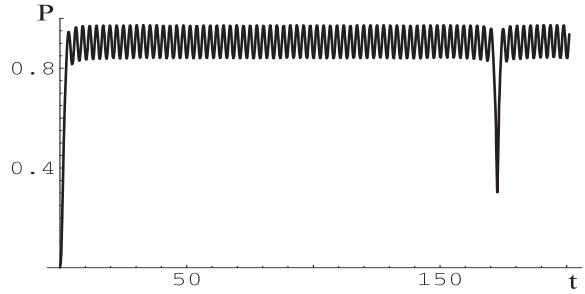}
\end{center}
\caption{The probability of population of the excited level as a result of a
two-fold transformation for different factorization constants: $f_{0}=1$, $%
R_{1}=0.993$, $R_{2}=0.995$, $\protect\xi =0.65$.}
\end{figure}
This behavior of probability continues for a long time (on Fig. 6 for
approximately $170$ time units that we use). However, the time during which
the probability of a transition to an exited state is less than $0.5$ (in
the present case it approximately is equal to $0.35$) is less than the
indicated interval by several orders of units (approximately two units in
the current case). This means that in the course of time the most part of
two-level atoms are found in the excited state, which may lead to the
appearance of an inverse population, as well as to a possible appearance of
the lasing property of an ensemble of two-level atoms placed in such kind of
field.

\section{Conclusion}

We have considered chains of SUSY transformations for the spin equation,
with an application to a two-level atom in an external electromagnetic field
with a possible time-dependence of the field frequency. It is shown that
when a chain of $n$ transformations is replaced by a single $n$-fold
transformation one can construct a super-Hamiltonian and supercharges which
close a polynomial superalgebra.
This fact permits us to state that a
polynomial pseudo-supersymmetry
can be associated with a
physical system, being a two-level atom in the present case.
It is discovered that, as
a result of a two-fold transformation, a certain type of time behavior of
the field frequency implies the disappearance of time oscillations of the
population probability for the excited level, and the probability becomes a
monotonously growing function of time with the limiting value which can
exceed $1/2$. This feature permits us to suppose that an ensemble of
two-level atoms placed in this specific electromagnetic field may acquire an
inverse population and exhibit lasing properties.

\section*{Acknowledgments}

The work is partially supported by grants SS-5103.2006.2 and
RFBR-06-02-16719. DMG is grateful to the foundations FAPESP and CNPq for
permanent support.


\begin{thebibliography}{99}
\bibitem{Nus73} H.M. Nussenzveig, Introduction to Quantum Optics, Gordon and
Breach, New York, 1973.

\bibitem{RabRaS54} I.I. Rabi, N.F. Ramsey, J. Schwinger, Rev. Mod. Phys.
26 (1945) 167.

\bibitem{FeyVe57} R.P. Feynman, F.L. Vernon, J. App. Phys. 28 (1957)
49.

\bibitem{Qcomp} Ya.S. Greenberg, Phys. Rev B 68 (2003) 224517.

\bibitem{Rabi} I.I. Rabi, Phys. Rev. 51 (1937) 652.

\bibitem{Levin} V.G. Bagrov, M.C. Baldiotti, D.M. Gitman, A.D. Levin, Ann.
Phys. 14 (2005) 764.

\bibitem{Bagrov} V.G. Bagrov, M.C. Baldiotti, D.M. Gitman, V.V.
Shamshutdinova, Ann. Phys. 14 (6) (2005) 390.

\bibitem{Shamshutdinova} B.F. Samsonov, V.V. Shamshutdinova, J. Phys. A 38
(2005) 4715.

\bibitem{CzJP} B.F. Samsonov, V.V. Shamshutdinova, D.M. Gitman, Czech. J.
Phys. 55 (9) (2005) 1173.

\bibitem{Witten} E. Witten, Nucl. Phys. B 188 (1981) 513;

Nucl. Phys. B 202 (1982) 253.

\bibitem{JPhys} I. Aref'eva, D.J. Fernandez, V. Hussin, J. Negro, L.M. Nieto, B.F. Samsonov, eds.,
Progress in Supersymmetric Quantum Mechanics (special issue of J. Phys. A 37 (2004) No 43).

\bibitem{Soliton} V.B. Matveev, in: P.C. Sabatier (Ed.), Problems Inverse,
Evolution nonlineare, CNRS, Paris, 1980;\newline V. Matveev, M.
Salle, Darboux transformation and solitons, Springer, New York,
1991;\newline S.B. Leble, M.A. Salle, A.V. Yurov, Inverse Problems
8 (1992) 207.

\bibitem{Cooper} F. Cooper, A. Khare, U. Sukhatme, Supersymmetry in quantum
mechanics, World Scientific, Singapore, 2001.

\bibitem{Nieto} L.M. Nieto, A.A. Pecheritsin, B. F. Samsonov, Ann. Phys.
(NY) 305 (2003) 151189.

\bibitem{Itzykson} C. Itzykson, J.-M. Drouffe, Statistical Field Theory,
vol.1, Cambridge University Press, Cambridge, 1989;\newline H.
Feshbach, Theoretical nuclear physics: nuclear reactions, Wiley,
New York, 1992.

\bibitem{Cannata} F. Cannata, M. Ioffe, R. Roychoudhury, P. Roy, Phys. Lett.
A 281 (2001) 305;\newline
F. Cannata, G. Junker, J. Trost, Phys. Lett. A 246 (1998) 219;\newline
A.A. Andrianov, M.V. Ioffe, F. Cannata, J.-P. Dedonder, Int. J. Mod. Phys. A
14 (1999) 2675;\newline
M. Znojil, Phys. Lett. A 259 (1999) 220;\newline
B. Bagchi, R. Roychoudhury, J. Phys. A 33 (2000) L1.

\bibitem{Bender1} C.M. Bender, S. Boettcher, Phys. Rev. Lett. 24 (1998) 5243.

\bibitem{QuasiH} F.G. Scholtz, H.B. Geyer, F.J.W. Hahne, Ann. Phys. (NY) 213
(1992) 74.

\bibitem{90302141} P.A.M. Dirac, Proc. Roy. Soc. London A 180 (1942) 1;%
\newline
W. Pauli, Rev. Mod. Phys. 15 (1943) 175;\newline
T.D. Lee, Phys. Rev. 95 (1954) 1329;\newline
S.N. Gupta, Phys. Rev. 77 (1950) 294;\newline
K. Bleuler, Helv. Phys. Act. 23 (1950) 567.

\bibitem{my2} B.F. Samsonov, J. Phys. A 38 (2005) L397.

\bibitem{my1} B.F. Samsonov, P. Roy, J. Phys. A 38 (2005) L249.

\bibitem{Crum} M.M. Crum, Quart. J. Math. 6 (1955) 121;\newline
M.G. Krein, Dokl. Akad. Nauk SSSR 113 (1957) 970.

\bibitem{Orszag} M. Orszag, Quantum optics, Springer-Verlag, Berlin, 2000.

\bibitem{Allen} L. Allen, J.H. Eberly, Optical Resonance and Two-level
Atoms, Wiley, New York, 1975.

\bibitem{Levitan} B.M. Levitan, Inverse Problems of Sturm-Leuvile, Nauka,
Moscow, 1984.

\bibitem{TMP} V.G. Bagrov, B.F. Samsonov, Theor. Math. Phys. 104 (1995) 1051.

\bibitem{Mostafazadeh} A. Mostafazadeh, J. Math. Phys. A 43 (2002) 205;

Nucl. Phys. B 640 (2002) 419.

\bibitem{Mielnik} B. Mielnik, O. Rosas-Ortiz, J. Phys. A 37 (2004) 10007.

\bibitem{Kamke} E. Kamke, Differentialgleichungen: L\"{o}sungsmethoden und L%
\"{o}sungen, 3-rd unaltered ed., Chelsea Pub. Co., New York, 1971.

\bibitem{JourPhys} B.F. Samsonov, A.A. Pecheritsin, J. Phys. A
37 (2004) 239.
\end{thebibliography}
\end{document}